\documentclass[twocolumn,aps,pre,floatfix,superscriptaddress,longbibliography]{revtex4-1}
\pdfoutput=1 
\usepackage{amsmath,amssymb,eucal,graphicx,float,epstopdf}

\usepackage[colorlinks=true, urlcolor=blue, anchorcolor=blue, citecolor=blue,filecolor=blue,linkcolor=blue,menucolor=blue]{hyperref}

\begin{document}
\title{Catalytic Coagulation}
\author{P. L. Krapivsky}
\affiliation{Department of Physics, Boston University, Boston, MA 02215, USA}
\affiliation{Santa Fe Institute, 1399 Hyde Park Road, Santa Fe, NM 87501, USA}
\author{S. Redner}
\affiliation{Santa Fe Institute, 1399 Hyde Park Road, Santa Fe, NM 87501, USA}

\begin{abstract}

  We introduce an autocatalytic aggregation model in which the rate at
  which two clusters merge is controlled by the third ``catalytic''
  cluster, whose mass must equal the mass of one of the reaction
  partners.  The catalyst is unaffected by the joining event and can
  participate in or catalyze subsequent reactions.  This model is
  meant to mimic the self-replicating reactions that occur in models
  for the origin of life.  We solve the kinetics of catalytic
  coagulation for the case of mass-independent reaction rates and show
  that the total cluster density decays as $t^{-1/3}$, while the
  density of clusters of fixed mass decays as $t^{-2/3}$.  These
  behaviors contrast with the corresponding $t^{-1}$ and $t^{-2}$
  scalings for classic aggregation.  We extend our model to
  mass-dependent reaction rates, to situations where only ``magic''
  mass clusters can catalyze reactions, and to include steady monomer
  input.
  
\end{abstract}  
\maketitle

\section{Introduction and Model}
\label{sec:no}

One of the profound mysteries of the natural world is the origin of
life.  Self-replication has been invoked as a starting point to
understand how the complex reactions that underlie living system might
arise, see, e.g.,
\cite{eigen1971selforganization,kauffman1971cellular,gilbert1986origin,Nowak08,Nowak09,Steen17,tkachenko2018onset,hordijk2019history,hilhorst}
and references therein. In such processes, the products of a given
reaction serve to catalyze the rate of new products, which, in turn,
can catalyze further reactions, leading to potentially to complex
chemistries.

Various types of random catalytic reaction networks have been proposed
and investigated to predict the emergence of autocatalytic cycles in
populations of diverse reactants with general types of catalytic
activity~\cite{stadler1993random,hanel2005phase,filisetti2012stochastic}.
The outcome of studies such as these is that catalytic activity among
a set of reactants is sufficient to promote the appearance of groups
of molecules that can replicate themselves through autocatalytic
reactions.

While the behavior of many of these autocatalytic reactions is
extremely rich, it is often not possible to discern which aspects of
the complex chemical reaction networks that have been studied are truly
necessary for the emergence of self-replication.  A missing element in
these models is analytical tractability---most of the models that have
been considered thus far typically contain many species and many
reaction pathways.  These complications make an analytical solution of
such models out of reach.  Motivated by this disconnect between
complexity and analytical tractability, we formulate a simple
realization of catalytic kinetics in the framework of irreversible
aggregation.  While our model is idealized, it might provide a starting point for analytically determining the kinetics of
autocatalytic reactions.

\begin{figure}[ht]
  \centerline{
    {\includegraphics[width=0.275\textwidth]{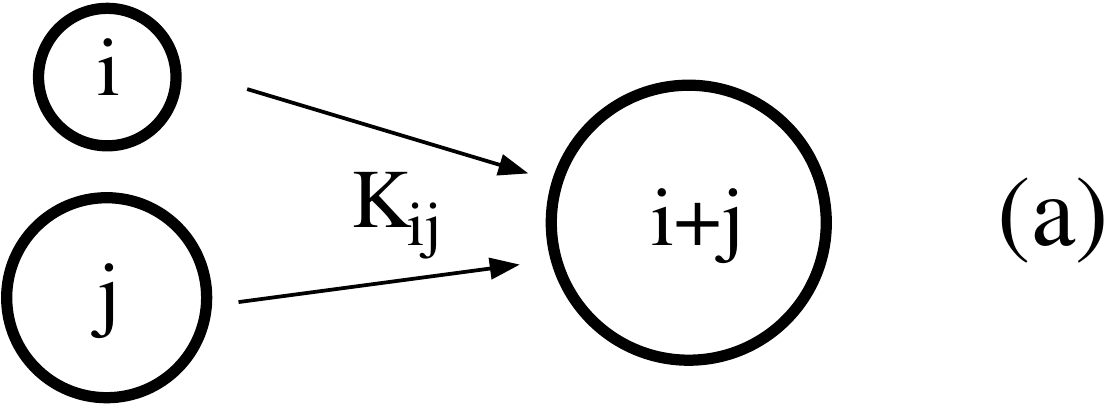}}}
\centerline{~~~}  
\centerline{~~~}  
\centerline{ {\includegraphics[width=0.275\textwidth]{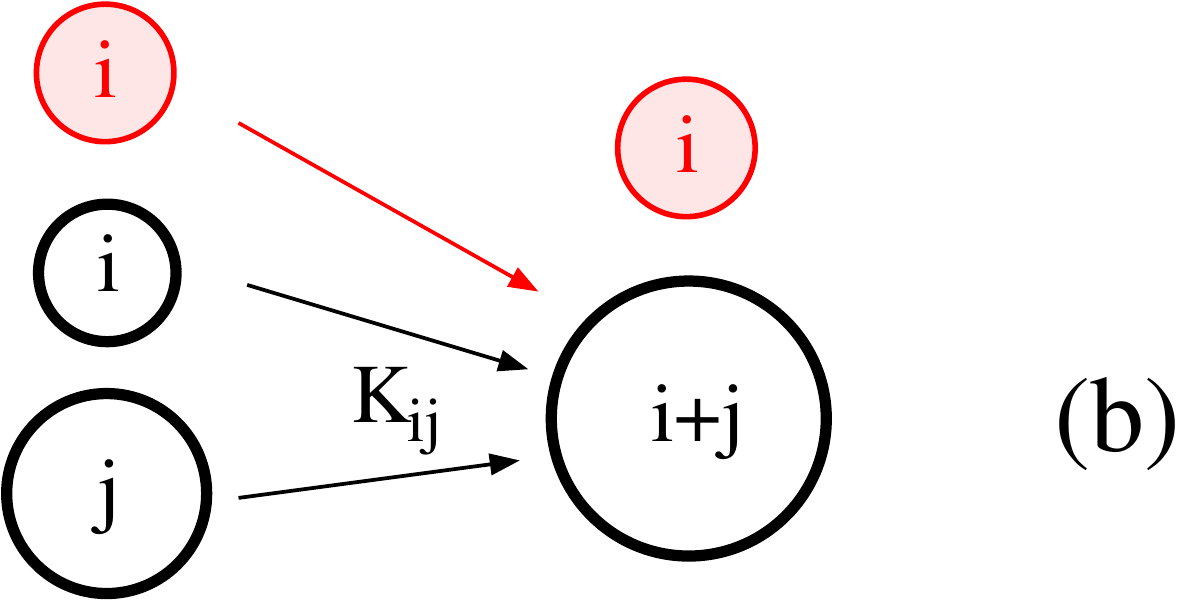}}}
  \caption{Schematic of the elemental events in: (a) coagulation and
    (b) catalytic coagulation.  In our catalytic coagulation model,
    the presence of a catalyst (red, shaded) whose mass matches one of
    the reactants is required.  This catalyst is unaffected by the
    reaction itself.}
  \label{fig:cartoon}
\end{figure}

In our \emph{catalytic coagulation} model, some fraction of the
reactions are catalytic; namely, these catalysts are unaffected by the
joining of two other reactants and can subsequently participate in or
catalyze further reactions.  That is, the rate at which a cluster of
mass $i$, an $i$-mer, and a $j$-mer join requires the presence of
either another $i$-mer or another $j$-mer to catalyze the reaction.
We may represent this reaction as
\begin{equation} 
\label{R:catalytic}
  \{i\} \oplus [i] \oplus [j]  \overset{K_{ij}}{\longrightarrow} \{i\} + [i + j]\,.
\end{equation}
Here, reactants inside square brackets undergo aggregation, while the
reactant within the braces is unaffected by the reaction. 

In the mean-field or perfect-mixing limit where all reactant
concentrations are spatially uniform, the catalytic aggregation
process \eqref{R:catalytic} has much slower kinetics compared to that
in conventional binary aggregation, $[i] \oplus [j] \to [i+j]$, and
even ternary (3-body) aggregation,
$[i] \oplus [j] \oplus [k] \to [i+j+k]$.  We may quantify this slower
kinetics by the temporal decay of the total cluster density $c(t)$
when the reaction rates are independent of the mass for all three
models.  This decay is given by
\begin{equation}
\label{c23-cat}
c(t)\sim 
\begin{cases}
t^{-1}   & \text{binary aggregation},\\
t^{-1/2}   & \text{ternary aggregation},\\
t^{-1/3}   & \text{catalytic aggregation}.
\end{cases}
\end{equation}
The $t^{-1}$ decay in binary aggregation immediately follows from the
closed equation that is satisfied by the total cluster density,
$\frac{dc}{dt}=-Kc^2$ (see Sec.~\ref{sec:binary}).  For ternary
aggregation, the decay of the concentration is described by
$\frac{dc}{dt}=-Kc^3$, from which $c\sim t^{-1/2}$ (see, e.g.,
\cite{PK91}).  The catalytic aggregation process \eqref{R:catalytic}
also involves a three-body interaction, but this interaction also must
satisfy the mass restriction that the catalyst mass matches the mass
of one of the two reactants.  This restriction is the source of the
slower decay compared to ternary aggregation.

In binary and ternary aggregation with mass-independent rates, we can
compute $c(t)$ in the entire time range.  In contrast, for catalytic
coagulation, the total cluster density does not satisfy a closed
equation even in the simplest case of mass-independent reaction rates.
Thus we are able to compute only the decay exponent,
$c(t)\sim t^{1/3}$ (Sec.~\ref{sec:constant}), but the amplitude
remains unknown.  We also derive the scaling solution of the
cluster-mass distribution.

In Sec.~\ref{sec:alg}, we extend our theory to treat the case where
the efficiency of the catalyst is a function of its mass.
Specifically, we analyze a one-parameter family of models with
algebraic reaction rates $K_{ij}=i^\nu$. In Sec.~\ref{sec:magic}, we
treat the situations where only clusters of certain ``magic'' masses
can catalyze reactions, as well as the situation where the reaction is
augmented by a steady monomer input.

\section{Classical Coagulation}
\label{sec:binary}

To set the stage for catalytic coagulation, we review some essential
features of classical coagulation.  Coagulation is a ubiquitous
kinetic process in which a population of clusters continuously merge
to form clusters of ever-increasing mass~\cite{Leyvraz03,KRB}.  This
process underlies many physical phenomena, such as blood clotting,
gravitational accretion of gas clouds into stars and planets, and
gelation.  In aggregation, two clusters of mass $i$ and $j$ join
irreversibly at rate $K_{ij}$ to form a cluster of mass $i+j$
according to
\begin{equation*} 
  [i] \oplus [j] \overset{K_{ij}}{\longrightarrow} [i + j]\,.
\end{equation*}
The basic observables are the densities of clusters of mass $k$ at
time $t$.  These $k$-mer densities depend in an essential way on the
reaction rates $K_{ij}$.  Much effort has been devoted to determining
these cluster densities in the perfectly-mixed or mean-field limit,
where the shape and spatial location of the clusters are ignored and
the only degree of freedom for each cluster is its
mass~\cite{Ziff,Ernst85a,Ernst85b,Ernst88}.

Let $c_k(t)$ denote the density of $k$-mers at time $t$.  In the
simplest aggregation process with mass-independent reaction rates, the
Smoluchowski equations \cite{Smo16,Smo17,Chandra} describing the
evolution of densities in the mean-field limit are particularly
simple:
\begin{equation}
\label{ck-eq}
\frac{dc_k}{dt}=\sum_{i+j=k}c_i c_j-2 c_k c
\end{equation}
where 
\begin{equation}     
\label{c:def}
c(t)\equiv  \sum_{k\geq  1} c_k(t)
\end{equation}
is the total cluster density. 
Summing Eqs.~\eqref{ck-eq} over all $k$, one finds that the total cluster density
satisfies $\frac{dc}{dt}=-c^2$, with solution $c(t)=(1+t)^{-1}$.

For the monodisperse initial condition $c_k(t\!=\!0)=\delta_{k,1}$, the
solution to \eqref{ck-eq} is
\begin{equation} 
\label{ck}
c_k(t)=\frac{t^{k-1}}{(1+t)^{k+1}}\,.
\end{equation}
In the scaling limit of $t\to\infty$ and $k\to\infty$, with the scaled mass 
$kc(t)$ kept finite, the mass distribution \eqref{ck} has the
scaling form
\begin{equation} 
\label{scaling-F}
c_k(t)\simeq c^2 F(ck)\,,
\end{equation}
with scaled mass distribution $F(x)=e^{-x}$.  We will compare these
classic results with the corresponding behavior of catalytic
coagulation in the following section.

\section{Catalytic Coagulation}
\label{sec:constant}

A variety of catalytic reaction schemes have been proposed and
investigated in the context building the complex molecules of living
systems~\cite{eigen1971selforganization,kauffman1971cellular,gilbert1986origin,Nowak08,Nowak09,Steen17,tkachenko2018onset,hordijk2019history,hilhorst}.
These models typically invoke some type of constraint in which the
size or composition of the catalyst matches, in some way, with the
reactants so as to facilitate a reaction.  For example,
Ref.~\cite{Steen17} proposed the catalytic reaction scheme
$\{i+j\}\oplus [i]\oplus [j] \to \{i+j\}\oplus [i+j]$, i.e., the
catalyst mass equals the sum of the two reactant masses.  By
construction, it is not possible to generate clusters whose masses
exceed the largest mass in the initial state.  Thus it is necessary to
augment this scheme with additional processes, as in~\cite{Steen17}, to have continuous
evolution.  The reaction process that we investigate
$\{i\} \oplus [i] \oplus [j] \longrightarrow \{i\} + [i + j]$, has the
advantage of leading to continuous evolution starting from the
monodisperse monomer-only initial condition without the need to invoke
additional reaction channels.

We initially assume that the rate of each of these reactions is
independent of the reactant masses and we set all reaction rates to 1.
The time evolution of the cluster densities now obey
\begin{align}
\label{ck-enzyme}
  \frac{dc_k}{dt}&= \frac{1}{2}\sum_{i+j=k}c_i c_j(c_i+c_j)- c_k\sum_{i\geq 1} c_i(c_i+c_k)\nonumber\\[1mm]
                  & =\sum_{i+j=k}c_i^2 c_j-c_k^2\,c  - c_k  Q\,,
\end{align}
which involves, in addition to the total cluster density
\eqref{c:def}, the quadratic moment of the mass distribution
\begin{equation}    
\label{Q:def}
Q(t) \equiv \sum_{k\geq  1} c_k(t)^2\,.
\end{equation}
Because the mass is manifestly conserved in each reaction, a useful
check of the correctness of the rate equations \eqref{ck-enzyme} is to
verify that $\sum_k k\frac{dc_k}{dt}=0$.

The presence of this quadratic moment renders the governing equations
\eqref{ck-enzyme} intractable.  To understand why, we recall that one
can solve the rate equations \eqref{ck-eq} for classical aggregation
recursively in terms of the known cluster density.  In catalytic
coagulation, the governing equations \eqref{ck-enzyme} are also
recurrent, but they require knowledge of both $c(t)$ \emph{and}
$Q(t)$.  Using Eqs.~\eqref{ck-enzyme}, these quantities obey
\begin{subequations}
\begin{align} 
\label{c-enzyme}
\frac{dc}{dt} &=-c Q,\\[2mm]
\label{Q-enzyme}
\frac{dQ}{dt}  &= 2\sum_{i\geq 1}\sum_{j\geq 1}c_i^2 c_j c_{i+j} -  2c\sum_{k\geq 1} c_k^3 - 2Q^2\,.
\end{align}
\end{subequations}
Equation \eqref{Q-enzyme} involves moments higher than quadratic, so Eqs.~\eqref{c-enzyme}--\eqref{Q-enzyme} do not
form a closed system and hence are not solvable.

As an alternative, we specialize to the long-time limit, where the
cluster mass distribution should have the scaling behavior
\eqref{scaling-F}.  We will see that Eqs.~\eqref{c-enzyme}--\eqref{Q-enzyme} can be
solved in this scaling limit.  For consistency with \eqref{c:def} and
with mass conservation, $\sum_{k\geq 1} k\,c_k=1$, the scaling function $F(x)$ must satisfy the conditions
\begin{equation} 
\label{F0}
\int_0^\infty dx\, F(x) = 1\quad\text{and} \quad
\int_0^\infty dx\, x\,F(x) = 1\,.
\end{equation}
By substituting the scaling form $c_k(t)\simeq c^2 F(ck)$ into
\eqref{Q:def} we obtain
\begin{equation}     
\label{Q-F}
Q = Ac^3, \qquad A =  \int_0^\infty dx\, F^2(x)\,.
\end{equation}
Finally, we substitute \eqref{Q-F} into \eqref{c-enzyme} and integrate
to obtain the cluster density in the long time limit:
\begin{equation}
\label{c-enzyme:sol}
c = (3 A t)^{-{1}/{3}}\,.
\end{equation}

Now that we have found the cluster density, let us determine the
monomer density.  Its governing equation is
\begin{equation}
\label{c1-enzyme}
\frac{dc_1}{dt}=-c_1^2c  - c_1Q=-c_1^2c  - Ac_1 c^3\,.
\end{equation}
Dividing \eqref{c1-enzyme} by $\frac{dc}{dt}=-cQ=-A c^4$ yields
\begin{equation}
\label{c1c}
 \frac{dc_1}{dc} = \frac{c_1^2+Ac_1c^2}{Ac^3}
\end{equation}
The behavior in classical aggregation, $c_1\simeq c^2$, suggests a
similar algebraic scaling, $c_1\simeq B c^\beta$, in catalytic
coagulation. Substituting this asymptotic into \eqref{c1c} gives
\begin{equation*}
(\beta-1) Bc^{\beta-1} \simeq \frac{B^2}{A}\,c^{2\beta-3}
\end{equation*}
Two possibilities emerge: $\beta=1$ when the left-hand side dominates,
and $\beta=2$ when both terms are of the same order, and we further
deduce $B=A$. A more accurate analysis based on substituting
$c_1\simeq Bc$ into \eqref{c1-enzyme} leads to inconsistent results,
and we thus conclude that
\begin{equation}
\label{c1c:sol}
c_1 = Ac^2\,.
\end{equation}
This equation for $c_1$ is consistent with the scaling form
\eqref{scaling-F} only if
\begin{equation} 
\label{FFF}
F(0)=A=\int_0^\infty dx\, F^2(x)\,,
\end{equation}

\begin{figure}[ht]
  \centerline{
    \includegraphics[width=0.5\textwidth]{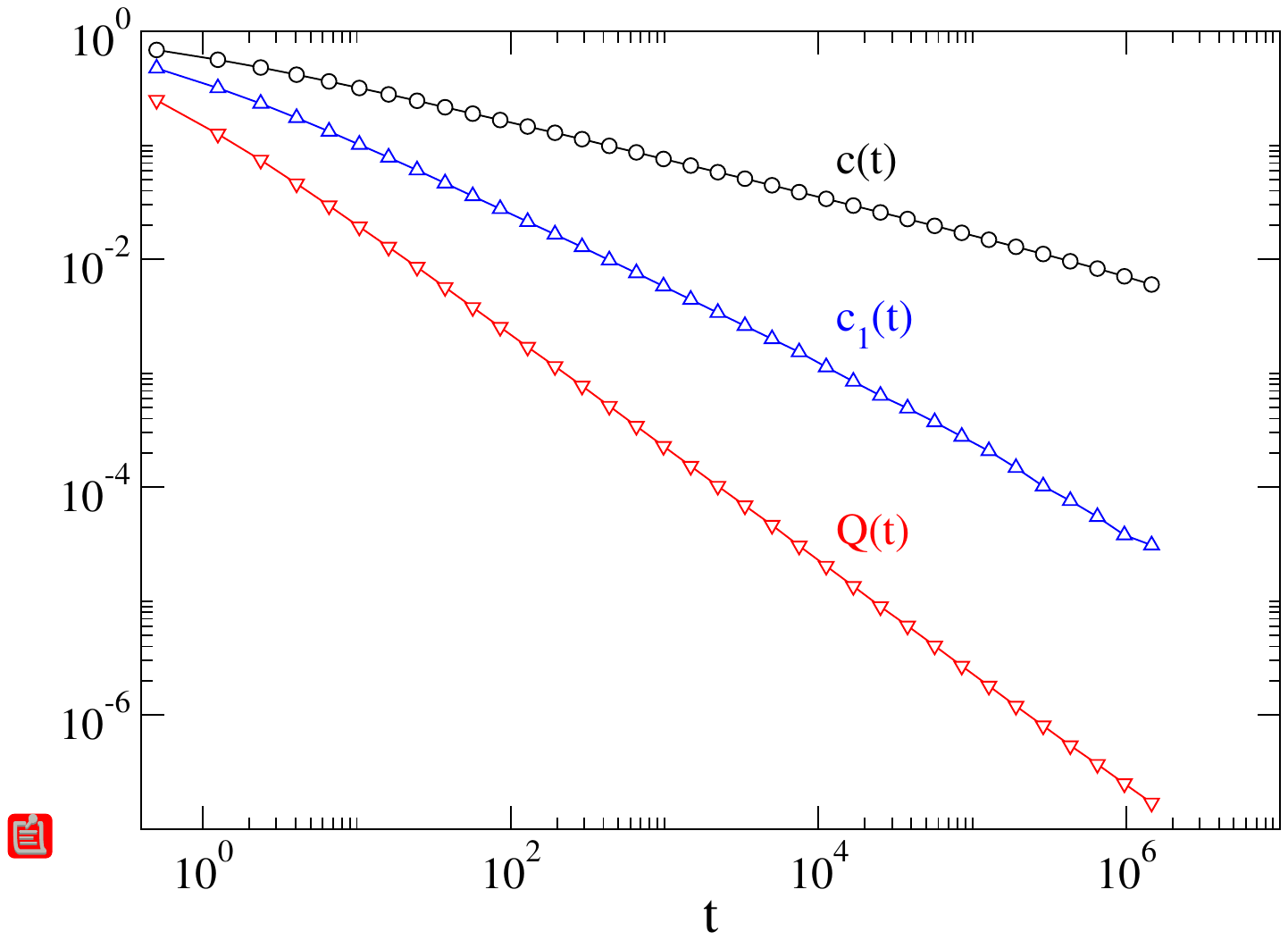}}
  \caption{Simulation data for $c(t)$, $c_1(t)$, and $Q(t)$ for
    catalytic coagulation on the complete graph of $10^5$ sites.}
  \label{fig:c-c1-q}
\end{figure}

Collecting \eqref{Q-F}, \eqref{c-enzyme:sol} and \eqref{c1c:sol} we arrive at
\begin{equation}
\label{cc-enzyme:sol}
c\simeq \frac{1}{(3At)^{1/3}}\, \quad c_1\simeq \frac{A^{1/3}}{(3 t)^{2/3}}\,,  \qquad Q \simeq \frac{1}{3t}\,.
\end{equation}
We have thus determined the asymptotic behavior of the quadratic
moment, while the densities of monomers and clusters are expressed in
terms of the unknown amplitude $A$.  As a check of these calculations,
Fig.~\ref{fig:c-c1-q} shows simulation data for $c(t)$, $c_1(t)$, and
$Q(t)$ in the mean-field limit for a $10^5$ realizations of the system
that initially contains $10^5$ monomers.  Least-squares fits to these
data on a double logarithmic scale in the time range
$10\leq t\leq 10^4$ give the respective slopes of $-0.327$, $-0.654$
and $-0.993$ compared to our predictions of $-1/3$, $-2/3$, and $-1$.
We also use the data to infer the amplitude $A$.  From
\eqref{cc-enzyme:sol}, the two combinations $Q/c^3$ and $c_1^3/Q^2$
should both approach $A$ for $t\to\infty$.  As a function of time,
both these variables converge to a common value up to $t\simeq 10^4$
before fluctuation effects begin to play a significant role. By this
analysis, we infer $A\approx 0.517$.

One point about the simulation worth mentioning is we absorbed the
factor $(c_i+c_j)$ in the right-hand side of Eq.~\eqref{ck-enzyme}
into the time, so that we again merely simulating binary aggregation,
but with a density dependent time increment.  This device makes the
simulation easy to code and quite efficient.

One can, in principle, continue this analysis to determine the $k$-mer
densities one by one.  However, it is more expedient to invoke
scaling.  Thus we substitute the scaling form \eqref{scaling-F} into
the rate equations \eqref{ck-enzyme}, from which we can directly
obtain the entire scaled mass distribution.  After some straightforward
algebra, the rate equations transform to the integro-differential
equation
\begin{subequations}
\begin{equation}
\label{F:long}
F^2-A\left[x\,\frac{dF}{dx}+ F\right] = \int_0^x dy\,F^2(y) F(x-y)\,.
\end{equation}
Notice that for $x=0$, the condition \eqref{FFF} that $F(0)=A$, is
automatically satisfied.

The transformation $\xi = Ax$ and $F(x) = A \Phi(\xi)$ recasts \eqref{F:long} into
\begin{equation}
\label{F:short}
\Phi^2-\Phi -\xi\,\Phi'=\int_0^\xi d\eta\,\Phi^2(\eta)\Phi(\xi-\eta)
\end{equation}
\end{subequations}
where the prime denotes differentiation with respect to $\xi$.  In
these new variables, Eq.~\eqref{FFF} becomes
\begin{equation}
\label{Phi-2}
1=\Phi(0)=\int_0^\infty d\xi\, \Phi^2(\xi)\,.
\end{equation}
The scaled mass distribution approaches to $\Phi(0)=1$ in the
small-mass limit of $\xi\to 0$.  To find the next correction, we write
$\Phi=1-\epsilon$ with $\epsilon\ll 1$, and substitute this ansatz
into \eqref{F:short} to find
\begin{equation}
\xi \epsilon' - \epsilon = \xi
\end{equation}
to leading order.  The solution is $\epsilon = \xi(\ln \xi + a)$, with $a$ 
some constant. Thus we conclude that the scaled mass distribution has the small-mass tail
\begin{equation}
\Phi = 1 -\xi(\ln \xi + a)+\ldots 
\end{equation}
as $\xi\to 0$. This small-mass behavior suggests that the scaled mass
distribution is more complicated than the scaled distribution
$\Phi=e^{-\xi}$ in classical aggregation.  While we have found the
small-mass tail of the scaled mass distribution, we have been unable
to determine the large-mass tail.

\section{Algebraic merging rates}
\label{sec:alg}

We can extend the approach of Sec.~\ref{sec:constant} to treat catalytic coagulation in
which the reaction rate depends on the mass of the catalyst: $K_{ij}=E_i$.  Such a
generalization accounts for the possibility that the efficacy of the
catalyst depends on geometrical constraints; for example, if the
catalyst serves as a physical scaffold upon which the reaction takes
place and a larger-area scaffold is more efficient.

We may write this mass-dependent catalytic reaction as
\begin{equation} 
\label{Ei}
\{i\} \oplus [i] \oplus [j]  \xrightarrow{\text{rate}~E_i} \{i\} \oplus [i + j]\,,
\end{equation}
in which the reaction rate $E_i$ is mass dependent.  A natural
situation is when the reaction rate is algebraic in the mass:
$E_i=i^\nu$.  On the physical grounds, the reaction rate cannot grow
faster than linearly in the mass, i.e., the exponent should satisfy
$\nu\leq 1$.  The $\nu>1$ range is not merely questionable physically,
but the resulting behaviors are often mathematically pathological.  In
an infinite system the process completes, that is, all clusters merge
into one, in zero time. This phenomenon of instantaneous gelation has
been studied in the context of classical aggregation, see, e.g.,
\cite{van87,BK91,Laurencot1999,Malyshkin01,Colm11}.  While
instantaneous gelation also seems to occur in catalytic coagulation,
we limit ourselves to the physically relevant range of $\nu\leq 1$.

For the catalytic reaction \eqref{Ei} with the reaction rate
$E_i=i^\nu$, the $k$-mer densities obey
\begin{equation}
\label{ck-nu}
\frac{dc_k}{dt}=\sum_{i+j=k}i^\nu c_i^2 c_j-kc_k^2\,c  - c_k Q_\nu\,,
\end{equation}
with 
\begin{equation}     
\label{Q-nu:def}
Q_\nu =\sum_{k\geq  1} k^\nu c_k^2\,.
\end{equation}

Similarly, the cluster density evolves according to
\begin{align} 
\label{c-nu}
\frac{dc}{dt} =-c\, Q_\nu\,.
\end{align}
When $\nu<1$, this mass=dependent catalytic coagulation admits a
scaling treatment parallel to that given in Sec.~\ref{sec:constant}
for the model with mass-independent rates ($\nu=0$). Combining
\eqref{Q-nu:def} with the scaling form \eqref{scaling-F} we obtain the
analog of Eq.~\eqref{Q-F}:
\begin{equation}     
\label{Q-nu-F}
Q_\nu = A_\nu c^{3-\nu}, \quad A_\nu =  \int_0^\infty dx\, x^\nu\,F^2(x)\,.
\end{equation}
Substituting \eqref{Q-nu-F} into \eqref{c-nu} and integrating, we obtain
\begin{equation}
\label{c-nu:sol}
c = [(3-\nu) A_\nu t]^{-1/(3-\nu)}
\end{equation}
for the density of clusters in the long-time limit.  Substituting \eqref{c-nu:sol} into
\eqref{Q-nu-F} we find the asymptotic behavior
\begin{equation}     
\label{Q-nu:sol}
Q_\nu = \frac{1}{(3-\nu)t}\,.
\end{equation}
Thus we know the exact asymptotic behavior of the moment $Q\nu$, while
the asymptotic of the more natural moment, the cluster density, is
solved only up to an unknown amplitude $A_\nu$.

The monomer density satisfies
\begin{equation}
\label{c1-nu}
\frac{dc_1}{dt}= -c_1^2\,c  - c_1\,Q_\nu\,.
\end{equation}
Dividing \eqref{c1-nu} by  \eqref{c-nu} and using $Q_\nu = A_\nu c^{3-\nu}$ we obtain
\begin{equation}
\label{c1c-nu}
\frac{dc_1}{dc}= \frac{c_1}{c}+\frac{c_1^2}{A_\nu c^{3-\nu}}\,.
\end{equation}
There are three possible alternatives for the asymptotic solution this
this equation: (i) The first term on the right-hand side of
\eqref{c1c-nu} is asymptotically dominant; (ii) the second term is
dominant; (iii) both terms are comparable. A straightforward analysis
shows that only the third possibility is consistent. Thus
$c_1\sim c^{2-\nu}$. Substituting this asymptotic into \eqref{c1c-nu}
we fix the amplitude:
\begin{equation}
\label{c1-nu:sol}
c_1 = (1-\nu) A_\nu c^{2-\nu}\,.
\end{equation}

Equation \eqref{c1-nu:sol} is compatible with the scaling prediction $c_1=c^2F(x)$ if
\begin{equation}
\label{F-nu-small}
F(x) = \frac{(1-\nu)A_\nu}{x^\nu}\qquad\text{as}\quad x\to 0\,.
\end{equation}

We can now obtain the governing equation for the scaled mass density
$F(x)$ by substituting the scaling form $c_k(t)\simeq c^2F(c k)$ into
\eqref{ck-nu} to give the analog of Eq.~\eqref{F:long}:
\begin{equation}
\label{F:nu}
x^\nu F^2-A\left[x\,\frac{dF}{dx}+ F\right] = \int_0^x dy\,y^\nu F^2(y) F(x-y)\,.
\end{equation}
As in the case of Eq.~\eqref{F:long}, the full equation is not analytically 
tractable, but it is possible to extract partial information about the
scaling functions in the limits of small- and large-$x$.

The asymptotic behaviors \eqref{c-nu:sol} and \eqref{Q-nu:sol} are valid for all $\nu\leq 1$, while \eqref{c1-nu:sol} is valid for $\nu<1$. A more careful analysis is required to establish the decay of the monomer density in the model with $\nu=1$, i.e.., with linear rates $E_i=i$. Specializing \eqref{c1c-nu} to $\nu=1$ we obtain
\begin{equation}
\label{c1c-1}
\frac{dc_1}{dc}= \frac{c_1}{c}+\frac{1}{A_1}\left(\frac{c_1}{c}\right)^2
\end{equation}
Substituting $c_1 = cu$ into \eqref{c1c-nu} gives
\begin{align*}
c\frac{du}{dc}=\frac{u^2}{A_1}\,,
\end{align*}
from which $u=A_1/\ln(1/c)$ when $c\to 0$. Using this together with
\eqref{c-nu:sol} and \eqref{Q-nu:sol} specialized to the case $\nu=1$
yields
\begin{equation}
Q_1=\frac{1}{2t}\,, \quad c=\frac{1}{\sqrt{2A_1 t}}\,, \quad c_1 = \sqrt{\frac{2A_1}{t}}\,\frac{1}{\ln(2A_1 t)}
\end{equation}
when $t\gg 1$.

\section{Catalysts with magic masses}
\label{sec:magic}

In many catalytic reactions, only a small subset of the reactants are catalytic.  Since the cluster mass is the only parameter in our modeling, the spectrum of masses for the catalytic reactants should be sparse, so that catalysts are rare. Here we treat an extreme model where only monomers are catalytic. In Appendix~\ref{ap:2n} we briefly consider the model where clusters with `magic' masses $2^n$ are catalytic.

\subsection{Only monomers are catalytic}
\label{subsec:mon}

If only monomers are catalytic, the reaction now is
$\{1\} \oplus [1] \oplus [j] \xrightarrow{\text{rate}~1} \{1\} \oplus
[1 + j]$.  The class of models \eqref{Ei} with algebraic reaction
rates $E_i=i^\nu$ reduces to the model where only monomers are
catalytic in the $\nu\to-\infty$ limit.

The cluster densities now evolve according to
\begin{subequations}
\label{c1k-mon}
\begin{align}
\label{c11}
\frac{dc_1}{dt} &=-c_1^2 (c+c_1)\,,\\
\label{c1k}
\frac{dc_k}{dt} &= c_1^2 (c_{k-1}- c_k), \quad k\geq 2\,.
\end{align}
\end{subequations}
Essentially the same equations describe the phenomenon of submonolayer
islanding~\cite{BK91,KRB}.  In the islanding reaction, monomers adsorb
and diffuse freely on a surface.  When two monomers meet or a monomer
meets a cluster of mass $k\geq 2$ merging takes place and all clusters
of mass $k\geq 2$ are immobile~\cite{BK91,KRB}.  The only difference
between submonolayer islanding and catalytic coagulation with
catalytic monomers is the factor $c_1^2$ instead of $c_1$ on the
right-hand sides of Eqs.~\eqref{c1k-mon}.

By introducing the modified time variable
\begin{equation}
\label{tau:def}
\tau = \int_0^t dt'\,c_1^2(t')\,,
\end{equation}
we linearize \eqref{c1k-mon} and obtain
\begin{align}
  \label{c1k-tau}
  \begin{split}
\frac{dc_1}{d\tau} &=-c- c_1\,,\\[2mm]
\frac{dc_k}{d\tau} &= c_{k-1}- c_k, \quad k\geq 2\,,\\[2mm]
\frac{dc}{d\tau} &=-c\,.
  \end{split}
\end{align}
The last equation is not independent, as it is obtained by summing the rate equations for
all the $c_k$.  Solving this last equation gives
\begin{subequations}
\label{ccc-tau}
\begin{equation}
\label{c-tau}
c(\tau) = e^{-\tau}\,.
\end{equation}
Then we solve the equation for $c_1$ and find
\begin{equation}
\label{c1-tau}
c_1(\tau)=(1-\tau)e^{-\tau}\,.
\end{equation}
Finally, we solve the rate equations for $c_k$ for $k\geq 2$
recursively and find
\begin{equation}
\label{ck-tau}
c_k(\tau) =\left(\frac{\tau^{k-1}}{(k-1)!}-\frac{\tau^k}{k!}\right)e^{-\tau}\,,
\end{equation}
\end{subequations}
for the monodisperse initial condition.

The time evolution ends at $\tau_\text{max}=1$, which corresponds to
$t=\infty$.  At this moment the density of monomers vanishes and the
reaction freezes.  The $k$-mer densities at this final time are
\begin{equation}
\label{ck-final}
c_k(t\!=\!\infty)=\frac{k-1}{k!}\,e^{-1}, \quad c(t\!=\!\infty)=e^{-1}
\end{equation}
While the dependence of the densities in Eqs.~\eqref{ccc-tau} in terms
of the modified time $\tau$ is the same as in submonolayer
islanding~\cite{BK91,KRB}, the dependence on the physical time is
different.  To determine the dependence on physical time, we exploit
the fact that the monomer density vanishes, $c_1(t)\to 0$, as
$t\to\infty$.  Using this fact, together with $c(\infty)=e^{-1}$, we
simplify \eqref{c11} to
\begin{align*}
  \frac{dc_1}{dt} \simeq-\frac{c_1^2}{e}\,,
\end{align*}
from which
\begin{equation}
\label{c1-asymp}
c_1(t)\simeq \frac{e}{t}\,.
\end{equation}
In contrast for submonolayer islanding, the density of monomers decays
exponentially with time, $c_1\sim e^{-t/e}$.  The asymptotic approach of all the other
$k$-mer densities to their final values is also algebraic:
\begin{equation}
\label{ck-asymp}
c_k(t)-c_k(\infty) \simeq -\frac{k^2-3k+1}{k!}\, \frac{e}{t}\,.
\end{equation}

While catalytic coagulation with an initial population of catalytic
monomers is solvable, it has the obvious limitation that all reactions
terminate in a finite time.  For this reaction with only monomers
being catalytic to continue ad infinitum, it is necessary to postulate
the existence of a source of monomers. This is the subject of the next
section.

\subsection{Input of monomers}
\label{subsec:input}

We now extend the model \eqref{c1k-mon} and postulate that monomeric
catalysts are injected at a constant rate. Indeed, in mimicking the
origin of life it is natural to consider open systems. Clusters can
spontaneously arise via external processes which we do not describe;
instead, we merely account for them as a steady input of
catalysts. One may anticipate that the balance between input and the
increase of mass due to aggregation manifests itself by driving the
system to a steady state. This steady-state behavior often arises in
classical aggregation (see e.g.,
\cite{Field65,hayakawa87,Colm12}). However, the outcome in the present
case is continuous evolution, as we now demonstrate.

With monomer input, we add the source to Eq.~\eqref{c11}:
\begin{align} 
\label{c1-mon-J}
\frac{dc_1}{dt} =-c_1^2 c- c_1^3 + J\,,
\end{align}
where $J$ is the source strength.  The $k$-mer densities with
$k\geq 2$ again satisfy Eqs.~\eqref{c1k}. The time
dependence of the density of non-catalytic clusters,
\begin{equation}
N = \sum_{k\geq  2}c_k\,,
\end{equation}
can be found by summing Eqs.~\eqref{c1k} for $k\geq 2$ and gives
\begin{subequations}
\begin{align} 
\label{N-mon}
\frac{dN}{dt} =c_1^3 \,.
\end{align}
It is also useful to rewrite \eqref{c1-mon-J} as
\begin{align} 
\label{c1-N-J}
\frac{dc_1}{dt} =-c_1^2 N- 2c_1^3 + J\,.
\end{align}
\end{subequations}

\begin{figure}[t]
\includegraphics[width=0.44\textwidth]{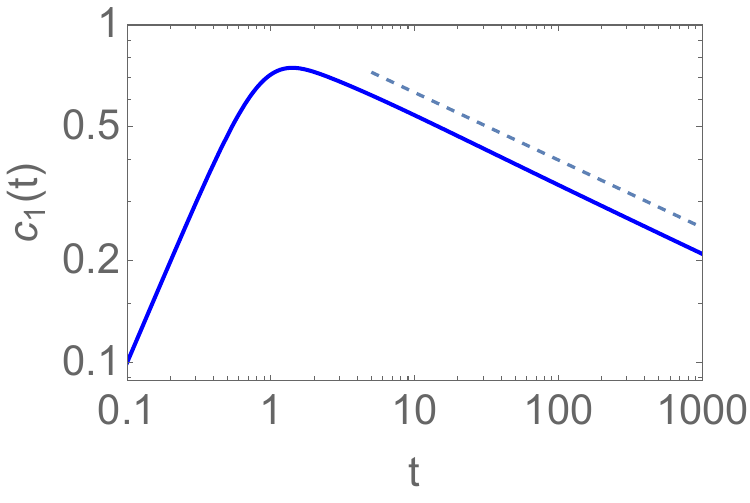}
\caption{The evolution of the monomer density from the numerical
  solution to Eqs.~\eqref{N-mon}--\eqref{c1-N-J} with $J=1$, subject
  to $c_1(0)=N(0)=0$.  The dashed line has slope $-1/5$.}
\label{fig:c1-mon}
\end{figure}

The pair of equations \eqref{N-mon}--\eqref{c1-N-J} do not admit an
exact solution but we can determine the asymptotic behavior. The
analysis is parallel to that given for the case of submonolayer island
growth~\cite{BK91,Blackman91,KRB}. It turns out (which can be
justified a posteriori) that $c_1\to 0$ and $N\to\infty$. Hence from
\eqref{c1-N-J} we obtain $c_1^2N\simeq J$, so that \eqref{N-mon} has
the asymptotic form $\frac{dN}{dt} =\left(\frac{J}{N}\right)^{3/2}$
leading to
\begin{align} 
\label{cN-mon}
  c\simeq  J^{3/5}\left(\frac{5t}{2}\right)^{2/5}, \qquad c_1 \simeq J^{1/5}\left(\frac{5t}{2}\right)^{-1/5}\,.
\end{align}
The numerical solution of Eqs.~\eqref{N-mon}--\eqref{c1-N-J} subject
to $c_1(0)=N(0)=0$ agrees with the asymptotic behaviors \eqref{cN-mon}
and it also demonstrate the initial development of the density of
monomers and the total cluster density.  The asymptotic behaviors
\eqref{cN-mon} imply that $c_1^2c\to J$ when $t\to \infty$, and this
is also readily confirmed by numerical integration of these equations.



By substituting $c_1\simeq (2J/5t)^{1/5}$ into \eqref{tau:def}, we
express the modified time in terms of the physical time and thereby express
the density in terms of the modified time 
\begin{align} 
\label{c1-mon}
c_1 \simeq  \left(\frac{2J}{3\tau}\right)^{1/3}\,.
\end{align}
Next we solve
\begin{align*}
  \frac{\partial c_k}{\partial \tau}= c_{k-1}- c_k\simeq
  -\frac{\partial c_k}{\partial k}\,.
\end{align*}
The solution to this wave equation is $c_k\simeq c_1(\tau-k)$.
Therefore
\begin{align} 
\label{ck-mon}
c_k \simeq \left(\frac{2J}{3(\tau-k)}\right)^{1/3}\,,
\end{align}
which is valid when $\tau-k\gg 1$. 

\section{Summary}

We introduced an aggregation process in which the reaction requires
the presence of catalysts.  These catalysts can both facilitate the
aggregation process and they can also directly participate in the
aggregation reactions.  While we vaguely have in mind the
self-replicating reactions that occur in models for the origin of
life, our modeling is more naive in character and is focused on
devising a set of reactions that both relies on catalytic action and
is analytically tractable.  By imposing the constraint that the mass
of the catalyst equals that of either of the participant in the
aggregation reaction, we have formulated a process that turns out to
be analytically tractable.  For the case where the the reaction starts
with a population of monomers, we solved the the $k$-mer densities in
the scaling limit and found that $c_k$ asymptotically decays as
$t^{-2/3}$, while the total cluster density decays as $t^{-1/3}$.  We
generalized our catalytic coagulation model to all for the efficacy of
the catalysts to grow with their mass.

We also extended our approach to deal with open systems.  This feature
of input of elemental reactants is a necessary ingredient to devise
models of artificial life.  In our modeling, we treated the situation
were only monomers are catalytic and they are injected into the system
in order to sustain a continuously evolving set of reactions.  For
this process, we can again solve for the kinetics of the reaction by
using classical tools of non-equilibrium statistical physics.  Here we
found that the cluster density continuously evolves, rather than
reaching a steady state, with $c(t)$ growing with time as $t^{2/5}$,
while the $k$-mer densities for fixed $k$ all decay with time as
$t^{-1/5}$.

\bigskip\noindent
We thank Steen Rasmussen and John Straub for helpful conversations.
This work has been partially supported by the National Science
Foundation under Grant No. DMR-1910736 and by the Santa Fe Institute.


%

\appendix
\section{Only clusters of mass $2^n$ are catalytic}
\label{ap:2n}

Here we study the process in which only reactants with `magic' masses
$2^n$ are catalytic. This leads to the following generalization of the
reaction \eqref{ck-enzyme}
\begin{equation*} 
\{2^i\} \oplus [2^i] \oplus [j]  \xrightarrow{\text{rate}~1} \{2^i\} \oplus [2^i + j]\,.
\end{equation*}

The time evolution of cluster densities with non-magic masses, $k\ne 2^n$, obey
\begin{subequations}
\begin{equation}
\label{ck-magic}
\frac{dc_k}{dt}=\sum_{2^i+j=k}b_i^2 c_j - c_k  Q\,,
\end{equation}
where $b_i\equiv c_{2^i}$ are the densities of clusters with magic
masses, while the density of magic-mass clusters obey
\begin{equation}
\label{bi-magic}
\frac{db_i}{dt}=\sum_{2^\ell+m=2^i}b_\ell^2 c_m-b_i^2 c  - b_i  Q\,.
\end{equation}
\end{subequations}
Here, we still denote the cluster density by $c$, while $Q$ is now the
quadratic moment of the mass distribution of magic-mass clusters
\begin{equation}     
\label{Q:magic}
Q = \sum_{n\geq  0} b_n^2\,.
\end{equation}
With this definition of catalytic clusters, the first of
Eqs.~\eqref{c-enzyme} still remains valid
\begin{align} 
\label{c-magic}
\frac{dc}{dt} =-c Q\,.
\end{align}

The validity of scaling is questionable. To appreciate this assertion,
consider the evolution of the densities of small-mass clusters. For
$k=1,2,3$ we obtain
\begin{align} 
\label{c1:magic}
  \begin{split}
    \frac{dc_1}{dt} &= -c_1^2c  - c_1  Q\,, \\
\frac{dc_2}{dt} &= c_1^3 -c_2^2c  - c_2  Q\,, \\
\frac{dc_3}{dt} &= c_1^2 c_2 + c_2^2c_1  - c_3  Q \,.
  \end{split}
\end{align}

The asymptotic behavior seemingly is 
\begin{align} 
  \label{c1:magic-a}
  \begin{split}
\frac{dc_1}{dt} &\simeq -c_1^2c \,,  \\
\frac{dc_2}{dt} &\simeq  -c_2^2c\,,  \\
\frac{dc_3}{dt} &\simeq  - c_3  Q \,.
  \end{split}
\end{align}
The decay of the densities of magic clusters $c_1$ and $c_2$ is
apparently qualitatively faster than the decay of $c_3$.  This
apparently different temporal behaviors for small-mass clusters
indicates that there no longer is a scaling description for the
cluster-mass distribution.

\section{Catalytic monomers with mass-dependent reaction rates}

The reaction scheme for this process is
\begin{equation*} 
\{1\} \oplus [1] \oplus [j]  \xrightarrow{\text{rate}~j} \{1\} \oplus [1 + j]\,.
\end{equation*}
In contrast to models we considered previously when the rate could
depend on the mass of the catalyst, we now assume that the rate
depends on the mass of the reactant.  The cluster densities now obey
\begin{subequations}
\label{c1k-linear-m}
\begin{align} 
\label{c1-linear-m}
\frac{dc_1}{dt} &=-c_1^2 (1+c_1)\,,\\
\label{ck-linear-m}
\frac{dc_k}{dt} &= c_1^2 [(k-1)c_{k-1}- kc_k], \quad k\geq 2\,.
\end{align}
\end{subequations}
In terms of the modified time \eqref{tau:def} we rewrite \eqref{c1k-linear-m} as 
\begin{subequations}
\begin{align} 
\label{c1-m-tau}
\frac{dc_1}{d\tau} &=-1- c_1\,,\\
\label{ck-m-tau}
\frac{dc_k}{d\tau} &= (k-1)c_{k-1}- kc_k, \quad k\geq 2\,.
\end{align}
\end{subequations}
Solving these equations recursively subject to the monodisperse initial condition yields \cite{BK91}
\begin{equation}
\label{ck-tau-linear}
c_k(\tau) = e^{-\tau}(1-e^{-\tau})^{k-1} - k^{-1}(1-e^{-\tau})^k\,.
\end{equation}
The density of monomers is
\begin{equation}
\label{c1-tau-linear}
c_1(\tau) = 2e^{-\tau} - 1\,,
\end{equation}
and it vanishes at $\tau_\text{max}=\ln 2$ corresponding to $t=\infty$. At this moment the process freezes. The final densities are
\begin{equation}
\label{ck-final-m}
c_k(t\!=\!\infty)=\frac{k-1}{k}\,2^{-k}, \quad c(t\!=\!\infty)=1-\ln 2\,.
\end{equation}
The latter formula follows from $c(\tau) = 1-\tau$ that follows by summing all the $k$-mer densities \eqref{ck-tau-linear}. 
In terms of the physical time, the asymptotic approach of the $k$-mer densities to their final values is algebraic. The leading behavior of  these corrections is inversely proportional to time
\begin{equation}
\label{ck-asymp-linear}
c_k(t)-c_k(\infty) \simeq -\frac{k-3}{2^k}\,t^{-1}\,.
\end{equation}
The only exception is the density of $3-$mers:
\begin{equation}
\label{c3-asymp}
c_3(t)-c_3(\infty) \simeq -\frac{1}{4 t^2}\,.
\end{equation}

We now inject catalysts with rate $J$ to counterbalance freezing. In
the system of equations \eqref{c1k-linear-m} only
Eq.~\eqref{c1-linear-m} is affected. In the case of initially empty
system we find
\begin{align} 
\label{c1-m-J}
\frac{dc_1}{dt} =-c_1^2(Jt+c_1) + J
\end{align}
Thus $c_1\simeq t^{-1/2}$ as $t\gg 1$. 

In the long time limit $c_k(t)$ approaches the scaling form 
\begin{subequations}
\label{scaling-J-m}
\begin{equation} 
\label{J-def-m}
c_k(t)\simeq t^{-1} F(k/t)
\end{equation}
More  precisely, this happens in the scaling limit
\begin{equation}
\label{scaling-m}
t\to\infty, \quad k\to\infty, \quad \frac{k}{t} = \text{finite}
\end{equation}
\end{subequations}
The scaled mass distribution can be extracted from the exact
formula in Ref.~\cite{BK91} (which is valid for all $k\geq 2$):
\begin{equation}
\label{ck-tau-exact}
c_k(\tau)=(k-1)\int_0^\tau du\,c_1(\tau-u)e^{-2u}\,\big[1-e^{-u}\big]^{k-2}
\end{equation}

Suppose $k=O(1)$. By substituting $c_1\simeq t^{-1/2}$ into \eqref{tau:def} and dropping terms that vanish as $t\to\infty$ we deduce that
\begin{equation}
\label{ttC}
\tau=\ln t + \ln C\,.
\end{equation}
(Fixing the constant $C$ requires an exact solution of \eqref{c1-m-J}
which looks intractable.) The asymptotic $c_1\simeq t^{-1/2}$ becomes
$c_1(\tau)\simeq \sqrt{C}\,e^{-\tau/2}$ in the modified time variable
$\tau$ when $\tau\gg 1$.  Substituting this latter form into
\eqref{ck-tau-exact} and using $U\equiv e^{-u}$ we deduce
\begin{eqnarray*}
c_k(\tau) &\simeq&  c_1(\tau) (k-1)\int_0^1 dU\,\sqrt{U}\,\big[1-U\big]^{k-2} \\
&=& c_1(\tau)\,\frac{\Gamma\big(\frac{3}{2}\big)\,\Gamma(k)}{\Gamma\big(k+\frac{1}{2}\big)}\,,
\end{eqnarray*}
in the long time limit. When $1\ll k\ll t$, we get
\begin{equation}
c_k(t)\simeq \sqrt{\frac{\pi}{4 kt}}\,,
\end{equation}
which is consistent with the scaling form \eqref{scaling-J-m} and gives the small-mass asymptotic of the scaled mass distribution
\begin{equation}
F(x)\simeq \sqrt{\frac{\pi x}{4}} \quad\text{when}\quad x\to 0\,.
\end{equation}

To extract the asymptotic behavior of the scaling function for
$x\gg 1$, we simplify the last factor in the integrand in
\eqref{ck-tau-exact}. Namely, we write $v=\tau-u$ and obtain
\begin{equation}
\label{ck-v}
c_k(\tau)\simeq (k-1)e^{-2\tau}\int_0^\tau dv\,c_1(v)\,\exp[2v-k  e^v e^{-\tau}]\,.
\end{equation}
Using \eqref{ttC} we find $k e^{-\tau}=x/C$. Hence in the exponent we
have $2v-k e^v e^{-\tau}=2v-e^v x/C$. Since $x\gg 1$, we only need the
small-$v$ behavior. We obtain
\begin{equation}
\label{ck-v-simple}
c_k(\tau) \simeq k\,e^{-2\tau-x/C}\int_0^\infty dv\,c_1(v)\,e^{-xv/C}\,.
\end{equation}
To compute the integral, we need to know the asymptotic behavior of
$c_1(v)$ when $v\ll 1$, as the integrand vanishes exponentially quickly when $v>1/x$. From
\eqref{c1-m-J} we find $c_1\simeq Jt$ when $t\ll 1$.
Substituting $c_1(t)\simeq Jt$ into \eqref{tau:def}
we obtain $\tau\simeq J^2t^3/3$, and hence
$c_1(\tau)\simeq (3J\tau)^{1/3}$, which recasts \eqref{ck-v-simple} into
\begin{eqnarray}
\label{ck-v-int}
c_k(t) &\simeq& \frac{k}{(Ct)^2}\,e^{-x/C}\int_0^\infty dv\,(3Jv)^{1/3}\,e^{-xv/C} \nonumber\\
&=& t^{-1}\,\left(\frac{3J}{C^2}\right)^{1/3}\,\Gamma\big(\tfrac{4}{3}\big)\,x^{-1/3}\,e^{-x/C}\,,
\end{eqnarray}
which is compatible with the scaling form \eqref{scaling-J-m} and gives the large mass asymptotic of the scaled mass distribution
\begin{equation}
F(x)\simeq \left(\frac{3J}{C^2}\right)^{1/3}\,\Gamma\big(\tfrac{4}{3}\big)\,x^{-1/3}\,e^{-x/C}\end{equation}
as $x\to \infty$. 

\end{document}